\documentclass[article]{revtex4}
\usepackage{amsmath}
\usepackage{amssymb}
\usepackage{graphicx}
\usepackage{indentfirst}
\usepackage{txfonts}
\usepackage{dcolumn}
\usepackage{bm}
\usepackage{epstopdf}
\usepackage{float}
\newcommand{\be}{\begin{equation}}
\newcommand{\ee}{\end{equation}}
\newcommand{\bea}{\begin{eqnarray}}
\newcommand{\eea}{\end{eqnarray}}
\newcommand{\ba}{\begin{eqnarray}}
\newcommand{\ea}{\end{eqnarray}}
\newcommand{\nn}{\nonumber}

\def\o{\omega}

\begin{document}
\title{Superradiant stability of dyonic black holes in string theory}
\author{Jia-Hui Huang, Mu-Zi Zhang, Tian-Tian Cao,Yi-Feng Zou}
\email{huangjh@m.scnu.edu.cn}
\affiliation{Guangdong Provincial Key Laboratory of Nuclear Science, Institute of quantum matter,\\
School of Physics and Telecommunication Engineering,
South China Normal University, Guangzhou 510006, China}
\author{Zhan-Feng Mai}
\email{zhanfeng.mai@gmail.com}
\affiliation{Center for Joint Quantum Studies,
School of Science, Tianjin University, Tianjin, China}
\date{\today}

\begin{abstract}
When a scalar wave perturbation is properly scattering off a charged or rotating black hole, the energy of the reflected scalar wave may be amplified. This is a superradiant process. If this amplification process can occur back and forth through certain confining mechanism, it will lead to strong instability of the black hole. In this paper, the superradiant stability is investigated for a special kind of dyonic black holes in string theory. Although the dynoic black hole has a similar spacetime metric with a electrically charged RN black hole, it is found that the dyonic black hole is more unstable than a RN black hole due to the coupling between magnetic charge of the dyonic black hole and the imping electrically charged scalar wave. We find two superradiantly stable regions in the parameter space  for the dyonic black holes and charged massive scalar perturbation.
\end{abstract}
\maketitle
\section{Introduction}
The (in)stability problem of black holes is an interesting
topic in black hole physics. Regge and Wheeler\cite{wheeler1957}
proved that the spherically symmetric Schwarzschild black hole is
stable under perturbations. The stability problems of rotating or
charged black holes are more complicated due to the significant
effect of superradiance. Superradiance effect can occur in both
classical and quantum scattering processes
\cite{Manogue1988,Greiner1985,Cardoso2004,Brito:2015oca,Brito:2014wla}. When a charged bosonic
wave is impinging upon a charged rotating black hole, the wave reflected by the
event horizon is amplified if the wave
frequency $\omega$ lies in the following superradiant regime
  \begin{equation}\label{suprad re}
   \omega < m\Omega  + e\Phi,
  \end{equation}
where $e$ is the charge of the bosonic wave, $m$ is the azimuthal number, $\Omega$ is the angular
velocity of black hole horizon and $\Phi$ is the electromagnetic
potential of the black hole
\cite{P1969,Ch1970,M1972,Ya1971,Bardeen1972,Bekenstein1973,Ya1972}. This means that when the incoming
wave is scattered, the wave extracts rotational energy from
rotating black holes and electromagnetic energy from charged
black holes. According to the black hole bomb mechanism proposed by Press and
Teukolsky\cite{PTbomb}, if there is a mirror between the black hole horizon and space
infinity, the amplified wave can be reflected back and forth
between the mirror and the black hole and grows exponentially.
This leads to the superradiant instability of the black hole.

The superradiant properties of different kinds of black holes have been extensively studied in the literature (for a review, see \cite{Brito:2015oca}).
A rotating Kerr black hole can be superradiantly unstable caused by  a massive scalar field when the parameters of the
black hole and the scalar filed are in certain  parameter spaces. In this case, the mass of the scalar field acts as a natural mirror. A lot of work has been done to identify these parameter spaces, e.g.
 \cite{Strafuss:2004qc,Konoplya:2006br,Cardoso:2011xi,Dolan:2012yt,Hod:2012zza,Hod:2014pza,Aliev:2014aba,Hod:2016iri,Degollado:2018ypf,Huang:2019xbu}.
A rapidly spinning black hole that is impinged upon by a complex and massive vector field is discussed in \cite{East:2017ovw,East:2017mrj}.
The superradiant instability of rotating black holes in curved space is also reported
\cite{Cardoso:2004hs,Cardoso:2013pza,Zhang:2014kna,Delice:2015zga,Aliev:2015wla,Wang:2015fgp,Ferreira:2017tnc,Konoplya:2013rxa,Kokkotas:2015uma}.

It is known that superradiant instability can also occur for charged black holes. When there is a mirror or a cavity outside a Reissner-Nordstrom (RN) black hole horizon, this black hole is superradiantly unstable in certain parameter spaces \cite{Herdeiro:2013pia,Li:2014gfg,Degollado:2013bha,Sanchis-Gual:2015lje,Fierro:2017fky,Gonzalez:2017gwa}. In flat background, RN black holes have been proved to be superradiantly stable against charged massive scalar perturbation\cite{Hod:2013eea,Huang:2015jza,Hod:2015hza,DiMenza:2014vpa}. It is pointed that when the parameters of a RN black hole and a charged massive scalar field satisfy superradiant conditions, there is no effective trapping potential outside the black hole horizon, which acts as a mirror to reflect the
superradiant modes \cite{Huang:2015jza}. When
charged black holes are in curved backgrounds, such as (anti-)de Sitter space, these backgrounds provide  natural reflecting boundary conditions and
support the existence of superradiant instability \cite{Wang:2014eha,Bosch:2016vcp,Huang:2016zoz,Gonzalez:2017shu,Zhu:2014sya}.
The discussion of black hole superradiant property has also been extended to the analogue of RN black hole in string theory  and
it has been shown that stringy RN black hole is  superradiantly stable against charged massive scalar perturbation\cite{Li:2013jna,Konoplya:2011qq}. But when  a
mirror is introduced, superradiant modes exist and the stringy RN black hole becomes unstable \cite{Li:2014xxa,Li:2014fna,Li:2015mqa}.

Magnetic field can trigger superradiant instability for a black hole. In \cite{Konoplya:2007yy,Konoplya:2008hj}, the authors use an approximation method to show that when a scalar field is propagating on the Ernst background, the magnetic field will induce an effective mass $\mu_{eff}\propto B$($B$ is the magnetic field strength) for the scalar field, leading to the superradiant instability. In a further work \cite{Brito:2014nja}, the authors considered Kerr-Newman black holes immersed in a uniform magnetic field and showed that  the magnetic field can confine scalar perturbations leading to long-lived modes, which trigger superradiant instabilities.

Besides electrically charged stringy RN black hole, another kind of interesting black holes in string theory are dyonic black holes \cite{Horowitz:1992jp,Li:2016nll,Sen:1992fr}, which have both electric and magnetic charges. In our previous work \cite{Huang:2015jza}, we showed that electrically charged  RN black holes are superradiantly stable under massive charged scalar perturbation. In this paper, we will consider the superradiant behavior of the dyonic black holes under a charged massive scalar perturbation. For our case, the metric of the dyonic black hole is similar to a
electrically charged RN black hole, however, the magnetic field also leads to the superradiant instability of the dyonic black holes and scalar perturbation system. In Section II, we describe the dyonic black hole and scalar perturbation system. In Section III, we analyse the radial equation of motion for the scalar and find two superradiant stable parameter regions for the system. Section IV is devoted to a summary.

\section{Dyonic black hole under a scalar perturbation}
In low energy effective theory of string
theory in four dimension, there are dyonic black hole solutions which
have both electric charge $Q_e$ and magnetic charge $Q_m$\cite{Horowitz:1992jp,Sen:1992fr}. It is convenient to use natural
units in which $G=c=1$. The
metric of the dyonic black hole in string theory is
\begin{equation}
d{s^2} =  - \frac{(r-r_+)(r-r_-)}{r^2-r^2_0}d{t^2}
+ \frac{r^2-r^2_0}{(r-r_+)(r-r_-)}d{r^2} + (r^2-r^2_0)(d\theta^2 +\sin^2\theta d\phi^2),
\end{equation}
where
 \bea
r_0=\frac{Q^2_m-Q^2_e}{2M},~~~r_{\pm}=M\pm(M^2+r^2_0-Q^2_m-Q^2_e)^{1/2}.
 \eea
$M$ is the mass of the black
hole. The dilaton filed $\phi$ is $e^{2\phi}=\frac{r+r_0}{r-r_0}$.
This black hole has an event horizon at $r_+$, inner horizon at
$r_-$ and a singular sphere at $r_0$.

In this paper, we consider a special dyonic black hole with equal electric and magnetic
charge $Q_e=Q_m=Q$.
Then, the above metric becomes to a RN-like  metric,
 \bea\label{metric}
 d{s^2} = - \frac{(r-r_+)(r-r_-)}{r^2}d{t^2}
+ \frac{r^2}{(r-r_+)(r-r_-)}d{r^2} + r^2 (d\theta^2+\sin^2 \theta d\phi^2),
 \eea
 where $r_\pm=M\pm(M^2-2Q^2)^{1/2}$. In this case, the dilaton filed $\phi=0$.
This metric has event horizon at $r_+$ and inner horizon at $r_-$ and there is
no singular sphere.
\emph{Although this metric is similar to RN black hole, we will see that its superradiance
behavior is different from RN black hole due to the magnetic charge.}

When a charged massive scalar perturbation $\Psi$ is impinging on the dyonic black hole,
the dynamics of the charged massive scalar field is described
by the Klein-Gordon (KG) equation
\begin{equation}\label{eom}
[({\nabla ^\nu } - iq{A^\nu })({\nabla _\nu } - iq{A_\nu }) - {\mu
^2}]\Psi  = 0,
\end{equation}
where $q$ and $\mu$ are the charge and mass of the scalar
field. The nonzero vector field components are $A_t=-Q/r$ and $A_\varphi=-Q \cos\theta$
which describe electric and magnetic fields of the black hole\cite{Sen:1992fr}.

The solution of the KG equation can be decomposed as the following form
\begin{equation}
\Psi(t,r,\theta,\phi)=R(r)Y(\theta)e^{im\varphi}e^{-i\omega t}.
\end{equation}
$ m$ is azimuthal harmonic index and $\omega$ is the frequency of the scalar perturbation.
Then, the angular part of the KG equation is
 \bea\label{angular equation}
 \frac{1}{\sin\theta}\partial_\theta(\sin\theta\partial_\theta Y)+
 (\lambda-\frac{m^2+q^2Q^2+2mqQ\cos\theta}{\sin^2\theta})Y=0.
 \eea
Substituting
$\eta=\cos\theta$
into the angular equation\label{angular equation}, one obtains the Fuchs equation
\be\label{Fuchs equation}
\frac{d^2Y}{d\eta^2}+(\frac{1}{\eta-1}+\frac{1}{\eta+1})\frac{dY}{d\eta}+[-\frac{(m+qQ)^2}{2(\eta-1)}+\frac{(m-qQ)^2}{2(\eta+1)}-\lambda]\frac{Y}{(\eta-1)(\eta+1)}=0,
\ee
which has three singularities ($-1,1,\infty$). The solutions of Fuchs equation \eqref{Fuchs equation} can be expressed by hypergeometric functions as
\bea
Y(\eta)=C(1-\eta)^{\frac{1}{2}(m+q Q)}(1+\eta)^{\frac{1}{2}(m-q Q)} {}_2F_{1}(\alpha, \beta; \gamma; \frac{1+\eta}{2}),
\eea
where $C$ is a normalization constant. The parameters $\alpha,\beta$ and $\gamma$  are given by
\begin{equation}
\left\{ \begin{array}{l}
{\alpha=m+ \frac{1}{2}(1+\sqrt{1+4\lambda})}\\
 {\beta=m+ \frac{1}{2}(1-\sqrt{1+4\lambda})}\\
 {\gamma=1+m-qQ}
 \end{array} \right.
\end{equation}
Since $Y(\eta)$ should be finite when $-1\leqslant\eta\leqslant1$, we have the following conclusions, \\
$\bullet$ $\beta$ is a negative integer, the separation constant $\lambda=l(l+1)$ and $l> m$.\\
$\bullet$ Finiteness of factor $(1-\eta)^{\frac{1}{2}(m-q Q)}$ in \eqref{Fuchs equation} implies  constraints between the  angular quantum numbers ($l$, $m$ ) of the scalar field and the charges of the scalar field and  black hole,
\be\label{mqQ}
m>q Q,~~l(l+1)>q^2 Q^2.
\ee
Different from the cases of RN black hole and the electrically charge black hole in string theory, due to the interaction between the charged scalar  and magnetic filed of the black hole, the solution of angular equation $Y(\eta)$ will be no longer hypergeometric function. Lower bounds on the angular quantum numbers of charged scalar modes appear.

The radial part of the Klein-Gordon equation is given by
\begin{equation}\label{ridial eq}
\Delta \frac{d}{dr}(\Delta \frac{dR}{dr})+UR=0,
\end{equation}
where $\Delta=(r-r_+)(r-r_-)$, and
\begin{equation}
U=(\omega r^2-qQr)^2-(r-r_+)(r-r_-)(-q^2 Q^2+\mu^2r^2+l(l+1)).
\end{equation}
In order to study the superradiance (in)stability of the black hole
against the massive charged perturbation, the asymptotic solutions
of the radial wave equation near the horizon and at infinity will be
considered with proper boundary conditions. Defining the tortoise
coordinate $y$ by the equation
\bea
\frac{dy}{dr}=\frac{r^2}{\Delta},
\eea
and a new radial function as ${\tilde R}=rR$, the
radial wave equation \eqref{ridial eq} can be written as
\begin{equation}
\frac{d^2\tilde R}{dy^2}+\tilde U\tilde R=0 ,
\end{equation}
where
\begin{equation}
\tilde
U=\frac{U}{r^4}+\frac{2\Delta^2}{r^6}-\frac{\Delta}{r^5}\frac{d\Delta}{dr}.
\end{equation}
The asymptotic forms of the new potential $\tilde U$ at the two
boundaries are
\begin{eqnarray}
\lim_{r \to {r_+}} \tilde U &=& (\omega-\frac{qQ}{r_+})^2,\\
\lim_{r \to +\infty } \tilde U &=& \omega^2 - {\mu ^2}.
\end{eqnarray}
For the existence of superradiance modes, the chosen boundary conditions
are ingoing wave at the horizon $(y\to -\infty)$ and bound states
(exponentially decaying modes) at spatial infinity $(y\to +\infty)$.
Then the radial wave equation has the following asymptotic solutions
\begin{equation}
 \tilde R\thicksim\left\{ \begin{array}{l}
{e^{-i(\omega-\frac{qQ}{r_+})y}},\quad \,\,y\to -\infty ~(r\to {r_+}) \\
 {e^{-\sqrt{\mu^2-\omega^2}y}},\quad \,\,y\to +\infty ~(r \to
 +\infty).\\
 \end{array} \right.
\end{equation}
It is obvious that $\tilde R$ describes exponentially decaying bound
state modes as $r \to
 +\infty$ when $\omega^2 < \mu^2$.

\section{stability analysis of dyonic black hole}
Now we analyze whether there is a trapping well outside the black
hole event horizon when the parameters of the
scalar field and the black hole satisfy the bound state condition
$\omega^2 < \mu^2$  and the superradiance conditionㄛ $0 < \omega < qQ/r_+$ \cite{raidancecondtion}. It is convenient to define a new radial function $\psi$ by
\begin{equation}
\psi=\Delta^{\frac{1}{2}}R.
\end{equation}
Using this new function, the radial equation (5) can be written in
the form of a Shrodinger-like wave equation
\begin{equation}
\frac{d^2\psi}{dr^2}+(\omega^2-V)\psi=0,
\end{equation}
where the effective potential is
\begin{equation}
V=\omega^2+\frac{1}{\Delta^2}(\Delta-U-(r-M)^2).
\end{equation}

In order to see if there exists a trapping well outside the event
horizon, we should analyze the shape of the effective potential $V$.
It is
obvious that
 \bea\label{voh}\nonumber
V \to -\infty,~~~ as~~~ r \to r_-;~~~~V \to -\infty,~~~ as~~~ r \to r_+;~~~~\\
V \to \mu^2+\frac{-4M\omega^2+2qQ\omega+2M\mu^2}{r}+O(\frac{1}{r^2}),~~~ as~~~ r \to +\infty.
 \eea
When $\omega$ satisfies superradiant condition, $\omega<\frac{qQ}{r_+}$, and bound state condition, $\omega<\mu$, we can prove
\bea\label{2plusroot}
-4M\omega^2+2qQ\omega+2M\mu^2>-4M\omega^2+2qQ\omega+2M\omega^2=2M\omega(\frac{qQ}{M}-\omega)>2M\omega(\frac{qQ}{r_+}-\omega)>0.
\eea
 This means the effective potential has at least one maximum in the region $r_-<r<r_+$ and  one maximum outside
the  event horizon ($r>r_+$). If there was only one maximum outside the event horizon for the
effective potential, there will be no trapping well and the black hole will be stable.
Defining a new radial variable $z=r-r_-$, we conclude that if the derivative of the effective potential,
$V'(z)$, has only two positive roots, the black hole will be superradiantly stable.

The derivative of the effective potential is
\begin{equation}
V'(z) =  \frac{-2}{\Delta^3}((\Delta-U-(z+r_- -M)^2)\frac{d\Delta}{dz}+\frac{1}{2}\Delta U').
\end{equation}
We just care about the roots of $V'(z)$, so that only the numerator of $V'(z)$ is needed for our discussion.
The numerator of $V'(z)$ can be written as a polynomial of $z$,
\begin{equation}\label{Vd}
f(z)=a z^4+b z^3+c z^2 +d z+e,
\end{equation}
where
\begin{eqnarray}
 {a} &=& -2M\mu^2+4M\omega^2-2qQ\o,\\
  {b} &=&4 (8 M^2 - 6 M r_+ + r_+^2)\omega^2-4 q Q (5 M - 2 r_+)\omega-2 (l + l^2 - 2 q^2 Q^2 + (6 M^2 - 6 M r_+ + r_+^2)\mu^2),\\
 {c} &=&12(2M-r_+)^3\omega^2-18qQ(2M-r_+)^2\omega-6[(M-r_+)l(l+1)-3Mq^2Q^2+2r_+q^2Q^2+\mu^2(M - r_+)(2 M-r_+)^2],\\
 {d}&=&4 (4 M - 3 r_+) (2 M - r_+)^3\omega^2-4 q Q (7 M - 5 r_+) (2 M - r_+)^2\omega-4 (2 M^2 - 3 M r_+ + r_+^2)^2\mu^2\\
 &&+4 (7 M^2 - 9 M r_+ + 3 r_+^2)q^2Q^2
 -4 (l(l+1)-1) (M - r_+)^2,\\
 {e} &=&2(r_--r_+)r^2_-(qQ-\o r_-)^2+\frac{1}{2}(r_--r_+)^3.
\end{eqnarray}

The four roots of $f(z)=0$ are denoted by $z_1, z_2, z_3, z_4$. They satisfy the following relations
 \bea\label{prod}
 z_1*z_2*z_3*z_4=e/a,\\\label{sum}
 z_1*z_2+z_1*z_3+z_1*z_4+z_2*z_3+z_2*z_4+z_3*z_4=c/a.
 \eea
Some comment on the roots is needed. In principle, these four roots are in complex plane. According to discussion below equation \eqref{2plusroot}, we already have two real roots. If there is a complex root, then its complex conjugate is also a root. In this case, the dyonic black hole is superradiantly stable. In the following discusstion, we suppose a worse case that  all four roots are real. 
From the superradiant conditions $\o < qQ/r_+$ and bound state condition $\o <\mu$, we can prove that
$a<0$ and $e<0$. From the asymptotic behaviors of the effective potential
analyzed before, we also know that there are at least two maxima for the effective potential $V(z)$ when $z>0$
and they correspond to two positive roots $z_1, z_2$ of $V'(z)=0$. The equation \eqref{prod} means the another two
roots $z_3, z_4$ are both positive or negative. If they are both negative, then there is no
trapping well outside the event horizon and the black hole is superradiantly stable. From equation \eqref{sum}, we
can find that $c/a <0$ (i.e. $c>0$) is a sufficient condition for negative $z_3, z_4$ and therefore a sufficient condition for
superradiantly stable dyonic black holes.

Now, let's  consider the parameter regions of dyonic black hole and scalar field where $c>0$. Coefficient $c$ can be treated as a quadratic polynomial of $\o$, $c=c(\o)$ and the coefficient of the quadratic term is positive. The intercept of $c(\o)$ is
\bea
c(0)=-6[(M-r_+)l(l+1)-3Mq^2Q^2+2r_+q^2Q^2+\mu^2(M - r_+)(2 M-r_+)^2].
\eea
The discriminant of the quadratic equation, $c(\o)=0$,  is
\bea
\Delta_c&&=\left(-18 q Q r_-^2\right)^2-4 \left(12 r_-^3\right) \left(3 \left(-r_- \left(l^2+l-3 q^2 Q^2-1\right)+r_+ \left(l^2+l-q^2 Q^2+1\right)-2 M-\mu^2 r_-^3+\mu^2 r_-^2 r_+\right)\right)\cr
&&=36 (2 M-r_+)^3 \left(8 l^2 (M-r_+)+8 l (M-r_+)+32 \mu ^2 M^3-64 \mu ^2 M^2 r_+-6 M q^2 Q^2+40 \mu ^2 M r_+^2+7 q^2 Q^2 r_+-8 \mu ^2 r_+^3\right).
\eea
In the following, we will discuss two cases for $c>0$ and find out the relevant parameter regions.

\subsubsection{ Case I:  $\Delta_c <0 \Rightarrow c > 0$}
By the definition, we have $2M > r_+>M$, and $\Delta_c <0 $ is equivalent to
\bea
8 l^2 (M-r_+)+8 l (M-r_+)+32 \mu ^2 M^3-64 \mu ^2 M^2 r_+-6 M q^2 Q^2+40 \mu ^2 M r_+^2+7 q^2 Q^2 r_+-8 \mu ^2 r_+^3<0.
\eea
The above inequality can be rewritten as
\bea\label{negDeltaC}
\mu^2>\frac{q^2Q^2}{r_+^2}\frac{x^2 ( 7 x-6)}{8 ( x-2)^2 ( x-1)}-\frac{l(l+1)}{r_-^2},
\eea
where $x=r_+/M$, and from the definition of $r_+$, we have $1<x<2$.  When the parameters of black hole and dyonic black hole satisfy Eq.\eqref{negDeltaC}, the dyonic black hole is superradiantly stable against the scalar perturbation. Two examples of the effective potential in this case are shown in Fig1.
\begin{figure}[H]
	\centering
	\includegraphics[width=0.7\linewidth]{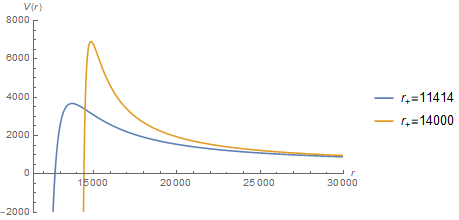}
	
	\caption{Two examples of the effective potential in case I. The black hole mass are chosen as
		$M=10500$ and $ M=10000$ for the blue curve and orange curve respectively. The other parameters are chosen as $l=\omega=5, \mu=20, q=15, Q=7000.$  }\label{Riemann1}
	
\end{figure}

\subsubsection{Case II: when $\Delta_c >0$, the two roots of $c(\omega)=0$ are $\omega_\pm$, then $0<\omega<\omega_- \Rightarrow c>0$}
The explicit forms of $\omega_\pm$ are
\bea
\omega_\pm=\frac{18qQr_-^2\pm\sqrt{\Delta_c}}{24 r_-^3}.
\eea
The condition $\Delta_c >0$ requires that $\mu$ must satisfy
\ba\label{deltaMu}
\mu^2<\frac{q^2Q^2}{r_+^2}\frac{x^2 ( 7 x-6)}{8 ( x-2)^2 ( x-1)}-\frac{l(l+1)}{r_-^2}.
\ea
We know that $\o$ satisfies superradiant condition, $\o < qQ/r_+$. If $qQ/r_+ <\omega_-$, then we have $0<\omega<\omega_- \Rightarrow c>0$.
So we consider the constraint $qQ/r_+ <\omega_-$,
\bea\nn
qQ/r_+ &<&\frac{18qQr_-^2-\sqrt{\Delta_c}}{24 r_-^3}=\frac{3qQ}{4r_-}-\frac{\sqrt{\Delta_c}}{24 r_-^3},\\
\Leftrightarrow\frac{\sqrt{\Delta_c}}{24 r_-^3}&<&\frac{3qQ}{4r_-}-qQ/r_+.
\eea
Then the above inequality is equivalent to
\bea
\frac{3qQ}{4r_-}-qQ/r_+>0~(r_+/r_->4/3),~~~ 18qQr_-^2r_+-24 r_-^3qQ>r_+\sqrt{\Delta_c}.
\eea
It is convenient to square the second inequality of the above so that we have
\ba\nn
r_+^2 (2 M - r_+)^2\mu^2-16 M^2 q^2 Q^2 + 20 M q^2 Q^2 r_+ + (l(l+1) - 7 q^2 Q^2) r_+^2>0,
\ea
which can be simplified as
\bea\label{poDeltaC}
\mu^2>\frac{q^2Q^2}{r_+^2}\frac{7r_+^2-20Mr_++16M^2}{r_-^2}-\frac{l(l+1)}{r_-^2}=\frac{q^2Q^2}{r_+^2}\frac{7x^2-20x+16}{(x-2)^2}-\frac{l(l+1)}{r_-^2}.
\eea
One can check that the above condition  is consistent with Eq.~\eqref{deltaMu},
\bea
\frac{q^2Q^2}{r_+^2}\frac{7x^2-20x+16}{(x-2)^2}-\frac{l(l+1)}{r_-^2}<\mu^2<\frac{q^2Q^2}{r_+^2}\frac{x^2 ( 7 x-6)}{8 ( x-2)^2 (x-1)}-\frac{l(l+1)}{r_-^2}.
\eea
So, together with the constraint on  mass and charge ratio of dyonic black hole, a superradiantly stable parameter region of the scalar and dyonic black hole system in case II is
\bea
\frac{q^2Q^2}{r_+^2}\frac{7x^2-20x+16}{(x-2)^2}-\frac{l(l+1)}{r_-^2}<\mu^2<\frac{q^2Q^2}{r_+^2}\frac{x^2 ( 7 x-6)}{8 ( x-2)^2 (x-1)}-\frac{l(l+1)}{r_-^2},\\
\frac{r_+}{r_-}>\frac{4}{3}~(\frac{Q}{M}<\frac{2\sqrt{6}}{7}).
\eea
There are two examples of the effective potential in this case are shown in Fig2.
\begin{figure}[H]
	\centering
	\includegraphics[width=0.7\linewidth]{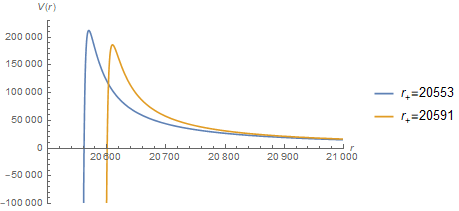}
	\caption{Two examples of the effective potential in case II. The black hole mass are chosen as $M=10000\sqrt{2}+1000$ and $ M=10000\sqrt{2}+1010$ for the blue curve and orange curve respectively. The other parameters are chosen as $l=5,\omega=7, \mu=15, q=14, Q=10000.$}
	\label{Riemann2}
\end{figure}

\subsubsection{Case III: when $\Delta_c >0$, the two roots of $c(\omega)=0$ are $\omega_\pm$, then $\omega>\omega_+ \Rightarrow c>0$}

We know that $\omega$ satisfies superradiant condition, $\omega < qQ/r_+$. In this case, we need  $\omega_+<\omega<qQ/r_+\Rightarrow c>0$.
Then we consider the constraint $qQ/r_+ >\omega_+$,
\bea\nn
qQ/r_+ &>&\frac{18qQr_-^2+\sqrt{\Delta_c}}{24 r_-^3}=\frac{3qQ}{4r_-}+\frac{\sqrt{\Delta_c}}{24 r_-^3},\\
\Leftrightarrow\frac{\sqrt{\Delta_c}}{24 r_-^3}&<&\frac{-3qQ}{4r_-}+qQ/r_+.
\eea
Then the above inequality is equivalent to
\bea
\frac{3qQ}{4r_-}-qQ/r_+<0~(r_+/r_-<4/3),~~~ -(18qQr_-^2r_+-24 r_-^3qQ)>r_+\sqrt{\Delta_c}.
\eea
It is convenient to square the second inequality of the above so that we have
\ba\nn
r_+^2 (2 M - r_+)^2\mu^2-16 M^2 q^2 Q^2 + 20 M q^2 Q^2 r_+ + (l(l+1) - 7 q^2 Q^2) r_+^2>0,
\ea
which can be simplified as
\bea\label{poDeltaC}
\mu^2>\frac{q^2Q^2}{r_+^2}\frac{7r_+^2-20Mr_++16M^2}{r_-^2}-\frac{l(l+1)}{r_-^2}=\frac{q^2Q^2}{r_+^2}\frac{7x^2-20x+16}{(x-2)^2}-\frac{l(l+1)}{r_-^2}.
\eea
One can check that the above condition  is consistent with Eq.~\eqref{deltaMu},
\bea
\frac{q^2Q^2}{r_+^2}\frac{7x^2-20x+16}{(x-2)^2}-\frac{l(l+1)}{r_-^2}<\mu^2<\frac{q^2Q^2}{r_+^2}\frac{x^2 (-6 + 7 x)}{8 (-2 + x)^2 (-1 + x)}-\frac{l(l+1)}{r_-^2}.
\eea

In case III, the superradiantly stable parameter space of the scalar and dyonic black hole system is
\bea
\frac{q^2Q^2}{r_+^2}\frac{7x^2-20x+16}{(x-2)^2}-\frac{l(l+1)}{r_-^2}<\mu^2<\frac{q^2Q^2}{r_+^2}\frac{x^2 (-6 + 7 x)}{8 (-2 + x)^2 (-1 + x)}-\frac{l(l+1)}{r_-^2},\\
\frac{r_+}{r_-}<\frac{4}{3}~(\frac{Q}{M}>\frac{2\sqrt{6}}{7}),\\
\omega_+<\omega<qQ/r_+.
\eea
\:Similarly,we draw two examples of the effective potential in this case are shown in Fig3.
\begin{figure}[H]
	\centering
	\includegraphics[width=0.7\linewidth]{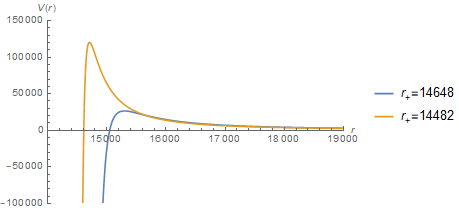}
	\caption{Two examples of the effective potential in case III. The black hole mass and frequency are chosen as $M=10000\sqrt{2}+10,\omega=6$ and $ M=10000\sqrt{2}+10,\omega=6.55$ for the blue curve and orange curve respectively. The other parameters are chosen as $l=5, \mu=15, q=10, Q=10000.$}
	\label{Riemann3}
\end{figure}

\section{Summary}
 In this paper, we study the superradiant stability of the dyonic black holes in string theory against a charged massive bosonic perturbation.
 Although the metric of a dyonic black hole is similar to that of a RN black hole, the magnetic field of the dyonic black hole makes the scalar and black hole system more unstable than RN black hole. The system is not superradiantly stale in whole parameter space and we discuss three cases for the superradiantly stable region.
For a general dyonic black hole (with the equal electric and magnetic charges) the superradiantly stable parameter region of the system is $\mu^2>\frac{q^2Q^2}{r_+^2}\frac{x^2 (7x-6 )}{8 (x-2)^2 (x-1)}-\frac{l(l+1)}{r_-^2}( x=r_+/M)$. If the ratio between the black hole charge and  the black hole mass satisfies $\frac{Q}{M}<\frac{2\sqrt{6}}{7}$, the superradiantly stable parameter region of the system becomes larger,
 which is $\mu^2 > \frac{q^2Q^2}{r_+^2}\frac{7x^2-20x+16}{(x-2)^2}-\frac{l(l+1)}{r_-^2}$. For the case $\frac{Q}{M}>\frac{2\sqrt{6}}{7}$, the superradiantly stable parameter is $\frac{q^2Q^2}{r_+^2}\frac{x^2 (-6 + 7 x)}{8 (-2 + x)^2 (-1 + x)}-\frac{l(l+1)}{r_-^2}>\mu^2 > \frac{q^2Q^2}{r_+^2}\frac{7x^2-20x+16}{(x-2)^2}-\frac{l(l+1)}{r_-^2}$  and $\omega_+<\omega<qQ/r_+$. For each case, we present a picture to show the superradiantly stable effective potential. 

Recently, it has been pointed out that the magnetically charged black holes have some interesting features\cite{Maldacena:2020skw}. They can have long-lived life even their masses are not large. They may have strong magnetic fields, which results in restoring the electroweak symmetry in some regions around them. It is worth noting that  a relevant interesting feature is the Hawking radiation effects are enhanced by the magnetic fields for near extremal black holes. In the near extremal limit, the Hawking emission modes are  in the superradiant region. In this sense, the magnetic fields enhanced the emission of superradiant modes. This is consistent with our result that magnetic fields make black hole system more unstable. It will be interesting to study further about magnetically charged black hole systems.

{\textbf{Acknowledgements:\\}}
Z.F.M thanks Professor H. L\"u and Shou-Long Li for useful discussion and proofreading. J.H.H. is supported by the Natural Science Foundation of Guangdong Province (No.2016A030313444). Z.F.M. is supported in part by NSCF grants No. 11475024 and No. 11875200.


\begin{thebibliography}{}
\bibitem{wheeler1957}
T. Regge and J. A. Wheeler, Phys. Rev. 108, 1063 (1957).
\bibitem{Manogue1988}
C.A. Manogue, Annals of Phys. 181 (1988) 261.
\bibitem{Greiner1985}
W. Greiner, B. M邦ller, J. Rafelski, Quantum Electrodynamics of
Strong Fields, Springer-Verlag, Berlin, 1985.



\bibitem{Cardoso2004}
  V.~Cardoso, O.~J.~C.~Dias, J.~P.~S.~Lemos and S.~Yoshida,
  Phys.\ Rev.\ D {\bf 70}, 044039 (2004)
  Erratum: [Phys.\ Rev.\ D {\bf 70}, 049903 (2004)]

 \bibitem{Brito:2015oca}
  R.~Brito, V.~Cardoso and P.~Pani,
  Lect.\ Notes Phys.\  {\bf 906}, pp.1 (2015).
  \bibitem{Brito:2014wla}
  R.~Brito, V.~Cardoso and P.~Pani,
  Class.\ Quant.\ Grav.\  {\bf 32}, no. 13, 134001 (2015).

\bibitem{P1969}
R. Penrose, Revista Del Nuovo Cimento,\textbf{1},252 (1969).
\bibitem{Ch1970}
D. Christodoulou, Phys. Rev. Lett. \textbf{25}, 1596 (1970).
\bibitem{M1972}
C. W. Misner, Phys. Rev. Lett. \textbf{28},994 (1972).
\bibitem{Ya1971}
Ya. B. Zel＊dovich, Pis＊ma Zh. Eksp. Teor. Fiz. 14, 270 (1971)
[JETP Lett. 14, 180 (1971)].
\bibitem{Ya1972}
Zh. Eksp. Teor. Fiz. 62, 2076 (1972)
[Sov. Phys. JETP 35, 1085 (1972)].
\bibitem{Bardeen1972}
J. M. Bardeen, W. H. Press, and S. A. Teukolsky, Astrophys. J. 178,
347 (1972).
\bibitem{Bekenstein1973}
J. D. Bekenstein, Phys. Rev. D 7, 949 (1973).
\bibitem{PTbomb}
W.H. Press, S.A. Teukolsky, Nature (London) 238, 211 (1972).
\bibitem{Brito:2015oca}
R.~Brito, V.~Cardoso and P.~Pani,
Lect. Notes Phys. \textbf{906}, pp.1-237 (2015).


\bibitem{Strafuss:2004qc}
  M.~J.~Strafuss and G.~Khanna,
  Phys.\ Rev.\ D {\bf 71}, 024034 (2005).
  \bibitem{Konoplya:2006br}
  R.~A.~Konoplya and A.~Zhidenko,
  Phys.\ Rev.\ D {\bf 73}, 124040 (2006).
  \bibitem{Cardoso:2011xi}
  V.~Cardoso, S.~Chakrabarti, P.~Pani, E.~Berti and L.~Gualtieri,
  Phys.\ Rev.\ Lett.\  {\bf 107}, 241101 (2011).
  \bibitem{Dolan:2012yt}
  S.~R.~Dolan,
  Phys.\ Rev.\ D {\bf 87}, no. 12, 124026 (2013).
  \bibitem{Hod:2012zza}
  S.~Hod,
  Phys.\ Lett.\ B {\bf 708}, 320 (2012).
  \bibitem{Hod:2014pza}
  S.~Hod,
  Phys.\ Lett.\ B {\bf 736}, 398 (2014).
  \bibitem{Aliev:2014aba}
  A.~N.~Aliev,
  JCAP {\bf 1411}, no. 11, 029 (2014).
  \bibitem{Hod:2016iri}
  S.~Hod,
  Phys.\ Lett.\ B {\bf 758}, 181 (2016).
    \bibitem{Degollado:2018ypf}
  J.~C.~Degollado, C.~A.~R.~Herdeiro and E.~Radu,
  Phys.\ Lett.\ B {\bf 781}, 651 (2018).
  \bibitem{Huang:2019xbu}
J.~Huang, W.~Chen, Z.~Huang and Z.~Mai,
Phys. Lett. B \textbf{798}, 135026 (2019).
  \bibitem{East:2017ovw}
  W.~E.~East and F.~Pretorius,
  Phys.\ Rev.\ Lett.\  {\bf 119}, no. 4, 041101 (2017).
  \bibitem{East:2017mrj}
  W.~E.~East,
  Phys.\ Rev.\ D {\bf 96}, no. 2, 024004 (2017).


\bibitem{Cardoso:2004hs}
  V.~Cardoso and O.~J.~C.~Dias,
  Phys.\ Rev.\ D {\bf 70}, 084011 (2004).
\bibitem{Cardoso:2013pza}
  V.~Cardoso, O.~J.~C.~Dias, G.~S.~Hartnett, L.~Lehner and J.~E.~Santos,
  JHEP {\bf 1404}, 183 (2014).
  \bibitem{Zhang:2014kna}
  C.~Y.~Zhang, S.~J.~Zhang and B.~Wang,
  JHEP {\bf 1408}, 011 (2014).
\bibitem{Delice:2015zga}
 O.~Delice and T.~Durgut,
  Phys.\ Rev.\ D {\bf 92}, no. 2, 024053 (2015)
\bibitem{Aliev:2015wla}
  A.~N.~Aliev,
  Eur.\ Phys.\ J.\ C {\bf 76}, no. 2, 58 (2016).
  \bibitem{Wang:2015fgp}
  M.~Wang and C.~Herdeiro,
  Phys.\ Rev.\ D {\bf 93}, no. 6, 064066 (2016).
  \bibitem{Ferreira:2017tnc}
  H.~R.~C.~Ferreira and C.~A.~R.~Herdeiro,
  Phys.\ Rev.\ D {\bf 97}, no. 8, 084003 (2018).

\bibitem{Konoplya:2013rxa}
  R.~A.~Konoplya and A.~Zhidenko,
  Phys.\ Rev.\ D {\bf 88}, 024054 (2013)

\bibitem{Kokkotas:2015uma}
K.~D.~Kokkotas, R.~A.~Konoplya and A.~Zhidenko,
  Phys.\ Rev.\ D {\bf 92}, no. 6, 064022 (2015)

\bibitem{Herdeiro:2013pia}
  C.~A.~R.~Herdeiro, J.~C.~Degollado and H.~F.~R迆narsson,
  Phys.\ Rev.\ D {\bf 88}, 063003 (2013).
  \bibitem{Degollado:2013bha}
  J.~C.~Degollado and C.~A.~R.~Herdeiro,
  Phys.\ Rev.\ D {\bf 89}, no. 6, 063005 (2014).
    \bibitem{Li:2014gfg}
  R.~Li, J.~K.~Zhao and Y.~M.~Zhang,
  Commun.\ Theor.\ Phys.\  {\bf 63}, no. 5, 569 (2015).
  \bibitem{Sanchis-Gual:2015lje}
  N.~Sanchis-Gual, J.~C.~Degollado, P.~J.~Montero, J.~A.~Font and C.~Herdeiro,
  Phys.\ Rev.\ Lett.\  {\bf 116}, no. 14, 141101 (2016).
  \bibitem{Fierro:2017fky}
  O.~Fierro, N.~Grandi and J.~Oliva,
  Class.\ Quant.\ Grav.\  {\bf 35}, no. 10, 105007 (2018).
\bibitem{Gonzalez:2017gwa}
  P.~A.~Gonz芍lez, R.~A.~Konoplya and Y.~V芍squez,
  Phys.\ Rev.\ D {\bf 95}, no. 12, 124012 (2017)






 \bibitem{Wang:2014eha}
  M.~Wang and C.~Herdeiro,
  Phys.\ Rev.\ D {\bf 89}, no. 8, 084062 (2014).
  \bibitem{Huang:2016zoz}
  Y.~Huang, D.~J.~Liu and X.~Z.~Li,
  Int.\ J.\ Mod.\ Phys.\ D {\bf 26}, no. 13, 1750141 (2017).
  \bibitem{Gonzalez:2017shu}
  P.~A.~Gonzalez, E.~Papantonopoulos, J.~Saavedra and Y.~Vasquez,
  Phys.\ Rev.\ D {\bf 95}, no. 6, 064046 (2017).


\bibitem{Hod:2013eea}
  S.~Hod,
  Phys.\ Lett.\ B {\bf 713}, 505 (2012).
  \bibitem{Huang:2015jza}
  J.~H.~Huang and Z.~F.~Mai,
  Eur.\ Phys.\ J.\ C {\bf 76}, no. 6, 314 (2016).
  \bibitem{Hod:2015hza}
  S.~Hod,
  Phys.\ Rev.\ D {\bf 91}, no. 4, 044047 (2015).
  \bibitem{DiMenza:2014vpa}
  Laurent Di Menza and Jean-Philippe Nicolas,
  Class.\ Quant.\ Grav.\  {\bf 32}, no. 14, 145013 (2015).

\bibitem{Wang:2014eha}
  M.~Wang and C.~Herdeiro,
  Phys.\ Rev.\ D {\bf 89}, no. 8, 084062 (2014).
    \bibitem{Bosch:2016vcp}
  P.~Bosch, S.~R.~Green and L.~Lehner,
  Phys.\ Rev.\ Lett.\  {\bf 116}, no. 14, 141102 (2016).
  \bibitem{Huang:2016zoz}
  Y.~Huang, D.~J.~Liu and X.~Z.~Li,
  Int.\ J.\ Mod.\ Phys.\ D {\bf 26}, no. 13, 1750141 (2017).
  \bibitem{Gonzalez:2017shu}
  P.~A.~Gonz芍lez, E.~Papantonopoulos, J.~Saavedra and Y.~V芍squez,
  Phys.\ Rev.\ D {\bf 95}, no. 6, 064046 (2017).
  \bibitem{Zhu:2014sya}
  Z.~Zhu, S.~J.~Zhang, C.~E.~Pellicer, B.~Wang and E.~Abdalla,
  Phys.\ Rev.\ D {\bf 90}, no. 4, 044042 (2014),
  Addendum: [Phys.\ Rev.\ D {\bf 90}, no. 4, 049904 (2014)].


\bibitem{Li:2013jna}
  R.~Li,
  Phys.\ Rev.\ D {\bf 88}, 127901 (2013).

\bibitem{Konoplya:2011qq}
R.~A.~Konoplya and A.~Zhidenko,
  Rev.\ Mod.\ Phys.\ {\bf 83}, 793 (2011)

\bibitem{Li:2014xxa}
  R.~Li and J.~Zhao,
  Eur.\ Phys.\ J.\ C {\bf 74}, no. 9, 3051 (2014).
\bibitem{Li:2014fna}
  R.~Li and J.~Zhao,
  Phys.\ Lett.\ B {\bf 740}, 317 (2015).
  \bibitem{Li:2015mqa}
  R.~Li, Y.~Tian, H.~b.~Zhang and J.~Zhao,
  Phys.\ Lett.\ B {\bf 750}, 520 (2015).




\bibitem{Konoplya:2008hj}
R.~Konoplya,
Phys. Lett. B \textbf{666}, 283-287 (2008).
\bibitem{Konoplya:2007yy}
R.~Konoplya and R.~Fontana,
Phys. Lett. B \textbf{659}, 375-379 (2008).
\bibitem{Brito:2014nja}
R.~Brito, V.~Cardoso and P.~Pani,
Phys. Rev. D \textbf{89}, no.10, 104045 (2014).



\bibitem{Horowitz:1992jp}
  G.~T.~Horowitz,
  [hep-th/9210119].

\bibitem{Li:2016nll}
  S.~Li, H.~L\"u and H.~Wei,
  JHEP {\bf 1607}, 004 (2016).
\bibitem{Sen:1992fr}
A.~Sen,
Nucl. Phys. B \textbf{404}, 109-126 (1993).


\bibitem{raidancecondtion}
Because the magnetic field does not do work on a charged particle, so we assume the superradiant condition for a dyonic black hole is
the same as that for a RN black hole.

\bibitem{Maldacena:2020skw}
J.~Maldacena,
[arXiv:2004.06084 [hep-th]].

\end{thebibliography}
\end{document}